# Current understanding of the processes underlying the triggering of and energy loss associated with type I ELMs

A. Kirk[1], D. Dunai[2] M. Dunne[3] G. Huijsmans[4], S. Pamela[1], M. Becoulet[5], J.R. Harrison[1],

J. Hillesheim[1], C. Roach[1], S. Saarelma[1]

[1]EURATOM/CCFE Fusion Association, Culham Science Centre, Abingdon, Oxon, OX14 3DB, UK
[2] KFKI RMKI, Association EURATOM, Budapest, Hungary
[3]Max-Planck Institut für Plasmaphysik, EURATOM Association, Garching, Germany
[4]ITER Organization, Cadarache, St. Paul-lez-Durance, France
[5]Association Euratom/CEA, CEA Cadarache, F-13108, St. Paul-lez-Durance, France

## Abstract

The type I ELMy H-mode is the baseline operating scenario for ITER. While it is known that the type I ELM ultimately results from the peeling-ballooning instability, there is growing experimental evidence that a mode grows up before the ELM crash that may modify the edge plasma, which then leads to the ELM event due to the peeling-ballooning mode. The triggered mode results in the release of a large number of particles and energy from the core plasma but the precise mechanism by which these losses occur is still not fully understood and hence makes predictions for future devices uncertain. Our current understanding of the processes that trigger type I ELMs and the size of the resultant energy loss are reviewed and compared to experimental data and ideas for further development are discussed.

## 1.    *Introduction*

The type I ELMy H-mode is the baseline operating scenario for ITER [1]. Extrapolations for the amount of energy released by Type I Edge Localised Modes (ELMs) in ITER, based on data from existing tokamaks, indicate that the largest ELMs could not be tolerated regularly because of the damage they would cause [2] and so finding a way of reducing the energy released is essential [3]. However, there is considerable uncertainty associated with such predictions because of the lack of understanding of all of the processes involved.

The type I ELM is thought to result from the peeling-ballooning MHD instability [4]. In this theory the edge pressure gradient grows in the inter-ELM period until the peeling-ballooning stability boundary is crossed at which point the ELM is triggered. However, as will be discussed in this paper it has been observed on several devices that the experimental profiles can exist near to this stability point for a substantial amount of the inter-ELM period raising the question: what ultimately triggers the ELM?

While the particle and energy losses from the plasma due to a type I ELM have been measured on a range of devices, including detailed measurements of the changes to the plasma profiles over the ELM crash, there is no detailed quantitative understanding of how these losses occur.

In this paper the current status of experimental measurements will be compared with results from models to review the current understanding of the processes that determine when a type I ELM is triggered and the resulting energy losses that occur. The layout of the paper is as follows: Section 2 describes the evolution of the edge plasma profiles during the inter-ELM period and examines the detailed changes that occur just before a type I ELM is triggered. In particular, it examines new evidence for a pre-cursor mode, attempts to identify it and discusses the role it may play in the ELM crash. Section 3 reviews the experimental evidence related to energy losses associated with type I ELMs and compares these observations with toy models and state of the art non-linear codes. Section 4 summarises the findings and discusses the direction for future studies.

## 2. *What triggers a type I ELM?*

### 2.1 Type I ELM cycle

There are two basic ideal MHD instabilities associated with ELMs: the ballooning mode and the peeling mode [5]. The peeling mode is associated with the edge current density, while the ballooning mode is driven by the pressure gradient. The often referred to picture for the evolution of the edge profiles between type I ELMs is that the pressure gradient increases on the transport time scales until it reaches the ballooning instability boundary, where further evolution of the gradient is clamped by high n (where n is the toroidal mode number) ballooning modes. The edge current then rises on resistive time scales, which were assumed to be slower, until the peeling-ballooning stability boundary is reached at which point a medium n instability is generated, the type I ELM. This leads to the loss of energy and particles which reduce the edge pressure and current until stability is reached. However, recent calculations have suggested that the edge pressure and current profiles evolve on very similar timescales during the inter-ELM cycle i.e. the edge pressure gradient is not established significantly in advance of the edge current.

Direct measurement of the edge current profile are still in their infancy [6][7]. On DIII-D the current profile during the inter-ELM period has been extracted using conditional averaging for multiple ELMs and does suggest some small but finite lag between the evolution of the pressure gradient and the edge current density [8]. Calculations of the current diffusion performed for ASDEX Upgrade discharges, show that the edge current density evolves on the same time scale as the temperature and density gradients [9][10]. The edge current density has also been reconstructed from magnetic measurements on ASDEX Upgrade using the CLISTE equilibrium code [11]. The evolution of the peak edge current density determined by this method versus the local peak edge pressure gradient is shown in Figure 1. An almost linear trend between the two quantities is observed throughout the ELM cycle and hence there is no evidence for a resistive delay in the build-up of the edge current density [12]. Similar results have been observed on DIII-D [13] where in this case the edge current density is calculated using the Sauter formula (see figure 8 in reference [14]).

In all of these cases the edge current and pedestal pressure gradient evolves to the peeling-ballooning stability boundary before the ELM is triggered but the ELM is not always triggered immediately when the stability boundary is reached/crossed. For example, on ASDEX Upgrade in some discharges the ELM is triggered immediately in others the profiles stop evolving and the pressure gradient remains stationary for a considerable fraction (up to 50 %) of the inter-ELM period before the ELM [10]. The local maximum pressure gradient being constant does not imply the distance to the stability limit is the same. In fact, on MAST it has been observed that although the pressure gradient remains fixed during most of the ELM cycle, the pressure pedestal height and width continue to growing [15]. The increasing width then reduces the stability limit such that it comes down to the experimental point shortly before the ELM.

In addition there are uncertainties in both the modelling and the experimental measurements but there is increasing evidence that the pedestal profiles exist in or near to the unstable region for a time before the ELM is triggered. This is particularly noticeable on JET, in a high refuelling scenario, where the pedestal profiles are stationary, near to the peeling-ballooning boundary, for 60 % of the ELM cycle [16]. This suggests that there must be some process that is stabilising the peeling-ballooning mode and, as will be discussed in the next section, one possibility for this is the edge velocity shear.

## 2.2 Growth of the peeling-ballooning mode

The Wilson and Cowley model [17] of the ELM predicts the explosive growth of the mode. The mode structure is elongated along a field line, localised in the flux surface, perpendicular to the field line and relatively extended radially. The ballooning mode grows explosively as the time approaches a "detonation" time when the theory predicts the explosive growth radially of narrow filaments of plasma, which push out from the core plasma into the Scrape Off Layer (SOL). Such filament structures have been observed experimentally, initially, using visible imaging on MAST [18] and subsequently using a variety of diagnostics on a range of devices (see [19] and references therein). The observation of filaments alone is not a proof of this model, however, because ballooning modes can also give rise to filaments. To prove the explosive part it would be necessary to show that the mode grows faster than exponential. At present there is no experimental evidence for or against an explosive growth and this should be the subject of future work.

An example of the observation of the toroidal/poloidal motion of the filament once they are beyond the LCFS is shown in Figure 2, which shows a wide angle view of ASDEX Upgrade plasma. Figure 2b shows an image obtained just after the start of the rise of the target $D_\alpha$ light associated with an ELM. Clear stripes are observed, which are on the outboard (low field side) edge of the plasma. In order to make the filaments more visible and to aid in analysis, a background subtraction is performed, which produces the image shown in Figure 2c. Figure 2d and e show the original and background subtracted images obtained 50 µs (5 frames) later where the filaments have rotated toroidally/poloidally in the co-current direction (or poloidally downwards in the ion diamagnetic direction) and 3 of the filaments have interacted with the ICRH limiter. The interactions show up as bright spots in the image. The toroidal mode number of these filaments is typically in the range 12-18.

In the Wilson Cowley model these filaments grow by twisting to push through magnetic field lines without reconnection and so would be impeded by the velocity shear that is known to exist in the pedestal region of the plasma (see for example [20][21]). Stability calculations have also suggested that the growth rate of the peeling-ballooning modes may be sensitive to the velocity and/or velocity shear in the pedestal region [22]. In fact in the presence of such a shear, it is difficult to see how the filament could squeeze between the field lines on neighbouring flux tubes i.e. they would most likely get distorted or even may be broken off as a blob of plasma.

The evolution of the toroidal rotation of the bulk plasma during an ELM has been measured at the edge of the plasma on DIII-D [20] and MAST [21]. Before the ELM there is a steep velocity gradient at the edge of the plasma. As the ELM progresses this shear is at first reduced and then removed completely before quickly recovering after the peak in $D_\alpha$ emission (see figure 1 in [20] and figure 13 in [21]). The start of the change in velocity occurs effectively at the same time as the $D_\alpha$ light starts to rise. There is little evidence that the change in velocity precedes the ELM, however, it should be noted that the integration time of these diagnostics are ~ 200 µs. Hence changes on time or spatial scales less than the resolution of the diagnostics may have been missed.

Although when the filaments have pushed out beyond the LCFS they clearly rotate in the co-current direction, there is the possibility that as the filaments push through the LCFS their poloidal rotation is small. Figure 3 shows measurements of the plasma edge during the ELM crash using a turbulence imaging beam emission spectroscopy (BES) diagnostic on MAST [23] which images a 8x16 cm region at the edge of the plasma with a 2cm spatial and 500 kHz temporal resolution. The first frame ($t_0$) shows the formation of the filament inside the LCFS. In the next 10 µs the filament extends out radially and then moves downwards poloidally in the co-current direction. In the next subsection how these filaments propagate though a region of high velocity shear will be discussed and in particular if there is any evidence for pre-cursor activity that could modify the edge plasma.

### 2.3 Evidence for pre-cursor activity

Evidence for pre-cursors with a high toroidal mode number that grow up before the crash of a type I ELM has been obtained in a variety of diagnostics on several devices. Fluctuations have been observed using ECE measurements on ASDEX Upgrade [24] and reflectometry on JT-60U [25]. These fluctuations are observed to grow up ~200 µs before the ELM and to be localised in a narrow region near to the top of the pedestal. On MAST density fluctuations have been observed in line average density measurements in the 100 μs before the filaments due to the ELM become visible at the Last Closed Flux Surface (LCFS) [26]. Pre-cursors have also been observed in magnetic sensors on JET [27] and in the coupling of ICRH into the plasma on both JET and AUG [28], which indicated that the perturbation rotated in the counter current (or electron diamagnetic) direction, although it was not clear if this rotation was for the pre-cursor of for the entire ELM event. Recently 2D ECE imaging performed on KSTAR [29] and ASDEX Upgrade [30] has measured the rotation of the pre-cursor activity. A pre-cursor mode is observed to grow ~ 200 μs before an ELM is triggered rotating in the electron diamagnetic (counter current) direction. The mode has a poloidal wavelength of 15 cm and radial size of 3cm, rotating with an apparent poloidal velocity of ~5 kms$^{-1}$ in the electron diamagnetic direction. The fluctuation frequency is ~20-50 kHz.

On MAST, a mode with similar characteristic has also been observed using the BES system [31]. Figure 4 shows the 2D images of the normalised density fluctuations separated in time by 5 μs. A clear upwards propagation of a mode structure can be observed. This upwards motion corresponds to a rotation in the counter current or electron diamagnetic direction. The mode is observed to start to grow about 100 μs before the ELM crash, with a frequency of ~20 kHz. It is located radially near to the top of the pedestal. The mode has a poloidal wavelength of ~ 10cm and a radial size of ~2cm. The toroidal mode number, n ~30-40. At the onset of the mode it is observed to rotate in the counter current direction while the background turbulence is observed to rotate in the co-current direction. The mode is observed to grow in size until it appears to lock the entire flow in the pedestal region (i.e. bring it to zero), at which point the filament structures associated with the ELM are observed to grow [31]. This mode has also been identified on MAST using Mirnov coils located on the low field side of the plasma [31] and using the Doppler back scattering (DBS) diagnostic. There is evidence that this mode may actually exist for a large proportion of the inter-ELM period but it fluctuates in size, remaining quite small until becoming strong ~ 100 μs before the ELM crash.

Gas puff imaging has been used to capture the two-dimensional evolution of type I ELM pre-cursors on NSTX [32]. This again shows that strong edge intensity modulations propagate in the electron diamagnetic direction while steadily drifting radially outwards. The intensity fluctuations were observed at frequencies around 20 kHz and wavenumbers of 0.05–0.2 cm$^{-1}$. Once in the SOL, the filaments reverse their propagation direction and travel in the ion diamagnetic direction. Edge intensity fluctuations are strongly correlated with magnetic signals from Mirnov coils.

In summary, recent observations on a wide variety of devices suggest that an electromagnetic pre-cursor mode located near to the pedestal top with a high toroidal mode number propagates in the electron diamagnetic direction and grows up rapidly just before a type I ELM crash. In the next subsection the possible identification of this mode will be discussed.

**2.4 Identification of the pre-cursor mode**

Recently there has been significant effort in performing gyro-kinetic simulations of the pedestal region. These studies have been motivated by an attempt to understand the microinstabilities that may ultimately determine the properties of the pedestal and in particular the pedestal width. The EPED model [33] proposes that drift wave turbulence is suppressed in the pedestal region by sheared flow and that turbulence associated with the kinetic ballooning mode (KBM) then constrains the pedestal to a critical normalized pressure gradient. The gyrokinetic code, GS2 [34], has been used for microstability analysis of the edge plasma region in MAST [15], JET [16] and NSTX [35]. In these simulations, while KBMs sometimes show up in the steep gradient pedestal region, Microtearing mode (MTM) are always found in the plateau region near to the top of the pedestal [36][37]. The MTM is an electromagnetic mode that is fundamentally driven by the electron temperature gradient and exists when the plasma normalised pressure ($\beta$) exceeds a critical value. The mode rotates in the electron diamagnetic direction, is stabilized by the density gradient and has a broadband frequency spectrum with poloidal wavenumber $0.6 < k_y \rho_i < 4$ [36], where $k_y$ is the perturbation wavenumber in the flux surface and perpendicular to magnetic field lines, and $\rho_i$ is the ion Larmor radius.

So could the experimentally observed pre-cursor mode be a MTM? While the location of the MTM at the top of the pedestal and the direction of rotation both agree with the experimental observations, the MTM is predicted to have a broadband frequency with frequencies above 200 kHz in the plasma frame. Correcting for the plasma rotation frequency, which is in the co-current direction, will reduce the frequency somewhat but the prediction of a broadband nature is harder to explain. However, to date the GS2 calculations performed are linear and it is possible that non-linear couplings may make the mode more localised and at a lower frequency.

Other candidates for the pre-cursor are the peeling mode but this would be expected to have a much lower toroidal mode number and be located nearer to the bottom of the pedestal i.e. closer to the separatrix. Drift waves are also expected in this region but the experimentally observed magnetic signature of the pre-cursors [31][32] rules out purely electrostatic drift waves. In addition, the fact that the pre-cursor is observed off mid-plane [30] is not expected for pure drift wave.

Of course the simplest would be if the pre-cursor was just the ballooning mode which then coupled to the peeling mode and resulted in the ELM. Then it would be necessary to explain how the same mode could change rotation direction. JOREK simulations shows that in the linear stage the ballooning mode perturbation grows and rotates in the electron diamagnetic direction and it is only in the non-linear stage that the perturbation changes to the ion diamagnetic direction as the filaments are expelled [38]. In the linear phase the large gradients cause a significant flow in the poloidal plane due to the finite resistivity. In the linear phase, the equilibrium flow is due to neoclassical and diamagnetic flow (and finite resistivity in older simulations) while in the non-linear phase the flow is driven by a non-linearity (Maxwell stress) in the momentum equation.

The only problem of identifying the pre-cursor with the linear stage of the ballooning mode is the mode number. In the JOREK simulations the ballooning mode either grows from low n to high or stays constant. This approach does not explain how a larger n pre-cursor can then become a lower n ELM. In reference [39], it has been suggested with that a radially localised high-n mode could destabilise a medium-n extended mode. The kinetic ballooning mode (KBM) could be a candidate for this high-n mode, however, the KBM is predicted to propagate in the ion diamagnetic direction and be localised in the steep gradient region, features not associated with the observed pre-cursor.

## 2.5 Different types of pre-cursors

While the high n pre-cursors described above have been seen on a variety of machines it is likely that other types of pre-cursors exist. For example, in addition to the pre-cursor observed in the ECE system and described above, two further distinct pre-cursors have been observed on ASDEX Upgrade. The first has been described as an "edge snake", which is located in the steep gradient region of the pedestal and appears to modulate the pedestal profiles [40]. The other is called the "Solitary magnetic perturbation" [41], which has also recently been observed on TCV [42]. Both of these pre-cursors propagate in the electron diamagnetic drift direction and so may again be providing a mechanism to reduce the edge flow shear and hence allow the peeling-ballooning mode to grow. In both cases the localisation of the instability leads to a dominant n=1 harmonic in the spectrum. The location of the perturbation near to the LCFS and its toroidal mode number may lead to the assumption that this could be the linear phase of the peeling mode. However, it could also be that this mode is still a ballooning instability at medium to high-n, that has become toroidally localised.

Such low n pre-cursors have been observed in simulations produced using the JOREK code [43]. These simulations were performed with a large set of toroidal harmonics and show that an n=1 harmonic, which would be weak in a linear simulation can achieve a large growth rate due to nonlinear coupling of higher dominant harmonics.

### 3. ELM energy loss mechanisms

The following is a possible interpretation of the temporal and spatial evolution of type I ELM: 1) the pre-cursor mode modifies the edge plasma leads to the formation of the peeling-ballooning mode that is at first localised inside the LCFS. 2) The peeling-ballooning mode grows producing filaments that push out beyond the LCFS. 3) The filaments separate completely from the plasma and travel radially to the wall losing particles by parallel transport. While this picture describes the stages it does not explain how the energy and particle loss associated with the ELM occurs.

It is known that the energy released by an ELM increases as the collisionality ($\nu^*$) is reduced [44]. At high collisionality the ELM energy loss is dominated by the change in density ($\Delta n$) while at low collisionality the changes in the temperature ($\Delta T$) pedestal become important. This has led to the losses being described as convective ($\Delta n$) and conductive ($\Delta T$) [45][46]. The convected component is effectively constant with collisionality but the conducted component increases strongly as the collisionality decreases [45][46]. As will be discussed below the mechanism for the ELM conducted energy loss, which dominates the losses at low $\nu^*$, is not understood. Therefore understanding this loss mechanism is essential in order to predict the total ELM energy loss.

As well as the pedestal affecting the ELM energy loss, results from JT60-U showed that the energy loss by a type I ELM is affected by the core pressure gradient inside the top of the pedestal [47]. The loss increases as the normalised pressure gradient inside the pedestal increases. This increase in ELM energy loss results directly from an increase in the drop of the pedestal ion temperature.

Describing these losses as conductive may not be correct and in fact may even be misleading. Since, if the losses were due to classical parallel conduction effects then since the electron conductivity ($\chi_e$) is much greater than the ion conductivity ($\chi_i$) then it would be expected that $\Delta T_e >> \Delta T_i$. However, this is not observed experimentally where $\Delta T_e \sim \Delta T_i$ [48]. As will be discussed later this may be too simplified a conclusion and in order to make quantitative predictions for ELM energy losses, models need to capture all the processes.

In the remainder of this section the key elements that may lead to ELM energy losses will be investigated, at first using toy models to discuss the role of filaments and edge ergodisation before looking at how the state of the art models correctly incorporate both effects into describing the entire ELM loss process.

### 3.1 The role of filaments in the energy loss process

Filament structures have been observed during ELMs on a number of devices [19] and these filaments have been shown to lead to the spiral patterns observed on the divertor [49][50]. Direct measurements have been made on MAST [50] and JET [51] and other devices of the energy content of the filaments. The maximum energy content of a single filament (assuming $T_i = T_e$), observed close to the LCFS on both devices, is ~2.5 % of $\Delta W_{ELM}$, but this may well be an under estimate since in the SOL $T_i > T_e$ [52]. The observations would suggest that the maximum amount of ELM energy transported by the separated filaments would be 25 % of $\Delta W_{ELM}$ ($T_i = T_e$) or 50 % of $\Delta W_{ELM}$ ($T_i = 3T_e$) for 10 equally sized filaments. Note that this assumes all the filaments have the maximum size observed and hence is likely an over estimate. Taking into account the spread in the measured energy content of the filaments and assuming that $T_i = T_i^{ped}$ (the ion temperature at the pedestal top), the total amount of energy carried away by the separated filaments is ~ 25-30% of $\Delta W_{ELM}$, which is in good agreement with the amount of ELM energy arriving at the first wall [53]. This means that 70-75 % of the ELM energy has to be lost, either by the filaments while they remain attached to surface inside the LCFS or by another process.

Filaments exist near the LCFS for 50 -100 μs at the start of the ELM event and are born as elongated field aligned structures, which have an initial parallel extent covering the low field side of the plasma. The rise time of the divertor ELM energy flux is correlated with the ion parallel transport time ($\tau_{//} = L_{//}/c_s$, where $L_{//}$ is the parallel connection length from the mid-plane to the divertor and $c_s$ is the ion sound speed) [54], and a detailed analysis of the temporal evolution of the ELM power target deposition in AUG and JET reveals that the ELM energy must be released from the core in $\leq$ 80 μs [55]. The correlation between this loss time and the time over which the filaments exist to the LCFS has led to the development of "leaky hosepipe" ELM energy loss models [56][50] where the filaments act as conduit for particles and energy from the confined plasma to the divertor. These models assume that the filament structures are linked to the core for a time duration of $\tau_{ELM}$. During this time the filament acts as a conduit for losses from the pedestal region into the SOL either by convective parallel transport due to a reconnection process (where one end of the filament remains attached to the core while other end becomes connected into the SOL) or by increasing the cross-field transport into the SOL. According to the predictions of non-linear ballooning mode theory, close to marginal stability $\tau_{ELM} \sim (\tau_A^2 \tau_E)^{1/3}$, where $\tau_A$ is the Alfvén time and $\tau_E$ is the energy confinement time [57]. The total number of particles that could flow down $n$ filaments is given by $N_{fil} = n\Gamma\sigma_{fil}\tau_{ELM}$, where $\Gamma$ is the average particle flux density ($\Gamma = \frac{1}{4}n\bar{c} = \frac{1}{4}n\sqrt{\frac{8kT}{\pi m_d}}$), ere $m_d$ is the mass of deuterium) and $\sigma_{fil}$ is the cross sectional area of the filament. The average particle flux density is calculated using the pedestal density and temperature respectively. The energy lost due to an ELM would then be given by the parallel convective heat equation i.e. $\Delta W_{ELM} = \frac{5}{2}k(T_i^{ped} + T_e^{ped})N_{fil}$. Assuming the filaments have a circular cross section, and using the measured widths, the ELM energy losses obtained from the model (assuming $T_i^{ped} = T_e^{ped}$) compared to the measured energy losses on MAST and JET are shown in Figure 5. The ELM energy loss model predictions are in reasonable agreement with the data given the crude nature of the model. However, what is not described is how the energy actually gets out of these filaments. Does one end of the filament reconnect into the LCFS while the other remains attached to the core, or does the perturbation to the field lines caused by the filaments increase cross field transport? Until the actual loss mechanism is established these models cannot be used to reliably extrapolate to future devices.

### 3.2 The role of ergodisation

In order to explain the ELM energy loss mechanism models based on the ergodisation of the edge plasma have been proposed where the ergodisation is driven by currents in the SOL or the filaments. As will be discussed in the next sub-section, the ergodisation does not need to be produced by current carrying filaments since it can be produced by the magnetic fluctuations associated with the ballooning mode itself. In the case of current carrying filaments, these currents are either thermoelectric in nature localised by intrinsic error fields [58][59] or due to the dynamo effect [60]. These models find that the plasma edge can be ergodised, and manifold structures form, when localised currents of 200-300 A are present. These values for the current are very similar to those observed experimentally in ELM filaments [61][62][63]. Models where the ELM energy loss is through ergodisation by current carrying filaments are attractive because they not only allow a mechanism by which the energy and particles can be removed from the core but also allow a mechanism for turning off the loss process as can be demonstrated with a toy model.

Consider starting off with 12 current carrying filaments located on the $q_{95}=5$ surface on MAST, where each filament carries 200 A (see Figure 6a). The effect that these filaments have on the edge plasma can be calculated, in the vacuum approximation using the ERGOS code [64], which is usually used to calculate the effect that externally applied resonant magnetic perturbations (RMPs) have on the plasma. In the case of current carrying filaments they are generated along field lines and hence the magnetic perturbations they produce are perfectly aligned with the equilibrium field. They therefore easily create island structures that overlap in the edge of the plasma (see Figure 7). In MAST experiments with externally applied RMPs there is a threshold in the normalised radial resonant field component ($b^r_{res}$) of the applied RMP field of $b^r_{res} \sim 0.5 \times 10^{-3}$, below which the applied RMPs have little effect on the plasma [65]. As can be seen in Figure 8 when the filaments are at the $q_{95}$ surface the peak value of $b^r_{res} = 4 \times 10^{-3}$ so a large effect would be expected on the plasma edge may be expected.

As the filaments separate and move away from the LCFS (Figure 6b and c) the maximum value of $b^r_{res}$ decreases rapidly and by the time the filament is 5cm from the LCFS the value of $b^r_{res}$ has approached the threshold value and hence the loss processes could be assumed to have stopped. These calculations have been done in the vacuum approximation and it is likely that if plasma screening is included the field would falloff much more quickly as the filaments move away from the LCFS.

ELM energy loss through ergodisation has certain implications that can be tested either now or in future experiments. Firstly, the splitting of the divertor strike point is a common observation in the presence of external non-axisymmetric magnetic perturbations (see for example [66][67][68][69]). Therefore edge ergodisation may be expected to produce a splitting of the strike point while the filaments are still inside the LCFS, and this splitting will be in addition to that produced once the filaments have separated. With ever faster cameras becoming available this is something that could be investigated in future experiments. Secondly the ergodisation would affect both the high and low field sides of the plasma and hence enhanced losses would occur to both the inner and outer divertor. However, in connected double null plasmas effectively no ELM energy goes to the high field side target (see for example [70][71][72]). But this may just point to our lack of understanding of how transport occurs in ergodic fields. For example, according to the Rechester–Rosenbluth model electron transport should dominate in the presence of an ergodic field and hence it may be expected that the dominant change would be to the electron temperature pedestal with little change to the ion temperature, which is not observed. The main consequence of applying RMPs to H-mode plasmas at low collisionality is the so called "density pump out" effect where the pedestal density drops while the temperature pedestal remains unaltered (see [73] and reference therein). In the case of RMPs this observation can be explained by changes in the radial electric field [74][75][76][77]. In particular it is found that due kinetic effects [77], the stochastic parallel thermal transport is significantly reduced compared to the prediction from the standard Rechester–Rosenbluth model [78]. The parallel electron heat transport is found to be approximately the same as the particle transport, which is significantly enhanced due to the changes in the radial electric field ($E_r$). The inclusion of such effects in modelling of the ELM should be a subject of future work.

### 3.3 State of the art modelling

Several non-linear codes have been developed in recent years to study the crash phase of ELMs. Considerable progress has been made and the codes are approaching realistic resistivities and are starting to see experimental phenomena and trends. Whilst some qualitative agreement has been achieved, the ability to accurately predict ELM sizes from first principles has not yet been demonstrated. Several nonlinear simulation codes have been applied to ELM simulations, including BOUT [79], BOUT++[80][81], GEM [82][83], JOREK [38][84][85], NIMROD [86][87] and M3D [88]. These employ different physical models, as well as a range of numerical methods. An unresolved issue is which set of equations is most relevant to ELM simulations, and what difference this makes to the predictions.

M3D [88], JOREK [38], BOUT++ and gyrofluid modelling [83] have reported ergodisation of the edge due to the magnetic fluctuations associated with the ballooning mode. For example, non-linear simulations of ELMs using the JOREK code [89] indicate that the footprint for the energy deposition pattern on the divertor is influenced by the ejection of filaments (which depends on their radial velocity) and heat conduction along homoclinic tangles (which depends on the size of magnetic perturbation) [90]. The lobe structures resulting from homoclinic tangles produced by the application of RMPs have been observed experimentally [91][92] . The challenge will be to see if similar structures can be observed during ELMs and disentangled from the effect that the filaments have in the X-point region.

These non-linear simulations have now reached a stage where they can be compared in detail with experimental data (see for example [84][85]). The simulated filament size and energy content are similar to the experimental observed ones [85] but the filaments are often more regularly spaced than in the experiment possibly due to the fact the simulations are often performed with only a single mode number. They are able to reproduce many of the divertor target foot print features and match well the experimentally observed broadening of the strike point with increasing ELM size [89]. However, to date they have not been able to match the in/out target heat load asymmetries observed on, for example, JET and ASDEX Upgrade [89]. In order to simulate the energy loss the codes often include a parallel conduction model (see for example [84]). While the incorporation of this enables the experimentally observed ELM energy loss scaling with collisionality to be recovered it does mean that the codes predict a larger change in the electron temperature compared to the ion temperature.     Figure 9 shows the fractional change in the electron and ion temperature profiles from before to after the ELM crash for the discharge modelled in reference [84]. Unlike the experimental observation, the change in the electron temperature profile is more than twice the change in ion temperature. This could be due to the choice of sheath boundary conditions chosen for this simulation. However, it should be noted that the change in ion temperature is not negligible because some of the changes in the ion temperature result from the expansion of the filaments into the SOL. 6-field simulations carried out with BOUT++ have shown cases where the ELM energy loss is mainly via the ion channel [93] but these simulations are for plasmas with a circular cross-section and it is not clear if they correctly capture the parallel loss processes.

In all these cases the non-linear simulations of ELMs start from an MHD unstable state i.e. the pedestal profiles have to be increased beyond the peeling-ballooning stability limit. The ELM size depends on the initial linear growth rate and hence how far above stability the simulation is started and this makes it difficult for them to reliably predict the ELM amplitude.

## *4. Summary and discussion*

In the inter-ELM stage of a type I ELM-ing discharge the pedestal pressure and current evolve on the transport time scale towards the peeling-ballooning stability limit. At some point the edge parameters cross this MHD stability boundary, the location of which can be reliably calculated using linear MHD codes. While ultimately it is the peeling-ballooning mode that leads to the type I ELM crash, the edge parameters can exist near to this unstable region for a substantial part of the inter-ELM period. There is growing experimental evidence that another mode grows up at the pedestal top that rotates in the electron diamagnetic direction and possibly reduces the edge flow shear sufficiently to allow the peeling-ballooning mode to grow explosively. Hence it would be appear that the flow shear at the pedestal top may be important for suppressing type I ELMs. In fact, exactly such a mechanism has been suggested as the reason that type I ELMs are suppressed in the QH-mode [94][95].

At least two types of pre-cursor have been observed one at high toroidal mode number (n~30-40) and one at low (n~1-2). In contrast to the peeling-ballooning mode which have intermediate toroidal mode numbers of n~10-15. The exact nature of the pre-cursor mode and the mechanism that triggers it is unclear, it could possibly be associated with micro tearing modes or else it could be a linear state of the peeling (for the low n) or ballooning (high n) modes. Ballooning mode simulations at very high Reynolds numbers ($S=2x10^8$) have shown the presence of pre-cursors before the ELM onset. The pre-cursor perturbation is located mostly inside separatrix and produces oscillation of ballooning mode at mid-plane, however, at present these simulations show that the toroidal mode number of the pre-cursor and peeling-ballooning mode are the same.

Filaments resulting from the peeling-ballooning mode clearly play some part in the ELM energy loss mechanism and it is likely that edge ergodisation also plays a role. At present the experimental measurements for this ergodisation are inconclusive. The non-linear models have now advanced to such a stage that they can now reproduce a lot of the experimentally observed phenomena. For example, they observe the formation of the filaments and the structure in the divertor heat flux profiles. But they have little predictive capability on the ELM amplitude. This is because most non-linear MHD simulations start from an MHD unstable state and the ELM size depends on the initial linear growth rate, which in terms depends on the distance above marginality. The inclusion of stabilisation by flows (diamagnetic, toroidal, poloidal) and how the loss of stabilisation occurs in the non-linear phase including the braking of flows due to MHD may be important to make progress in this area. It should be noted that the explosive growth of the ELM, which allows fast growth even close to marginal stability has yet to be observed in non-linear MHD simulations but as discussed earlier neither has it been observed experimentally.

If the trigger for the ELM crash can be determined, the next question is what are the loss mechanisms and what determines the final post ELM pedestal state, which is well below the marginal stability limit? At present the models suggest two loss processes one based on the filaments and the other based on ergodisation. Filaments arise due to the formation of ballooning interchange convective cells which move across the separatrix. There is parallel convective and conductive transport to divertor while the filament moves towards the wall. The magnetic perturbations due to the ballooning mode create homoclinic magnetic tangles, which give a direct connection from inside the pedestal to the divertor and parallel conductive losses occur with a strong temperature dependence. However, these loss mechanisms still do not capture the observed changes in the ion and electron temperature profiles and it may be necessary to include kinetic effects in the model. This may be possible by coupling the MHD codes with kinetic codes as was reported in the coupling of the XGC0 code with M3D [96].

### Acknowledgement

This work was part-funded by the RCUK Energy Programme [grant number EP/I501045] and the European Communities under the contract of Association between EURATOM and CCFE. To obtain further information on the data and models underlying this paper please contact PublicationsManager@ccfe.ac.uk. The views and opinions expressed herein do not necessarily reflect those of the European Commission

**Figures**

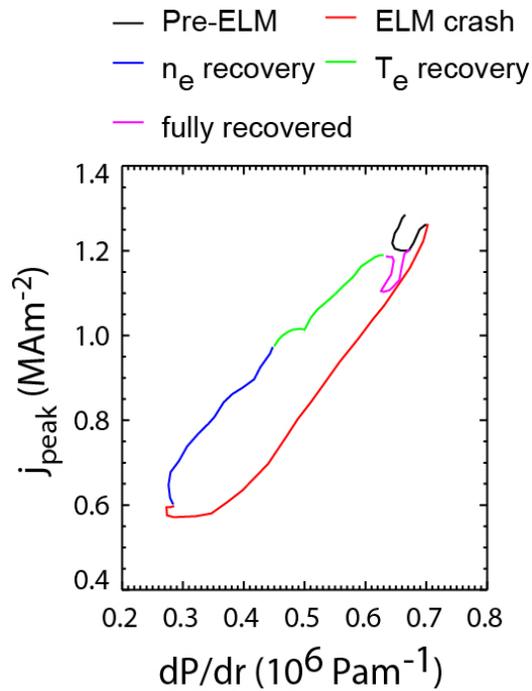

**Figure 1** The peak edge current (j$_{peak}$) vs the radial pressure gradient (dP/dr) calculated using the CLISTE equilibrium code as a function of time during an inter-ELM period on ASDEX Upgrade.

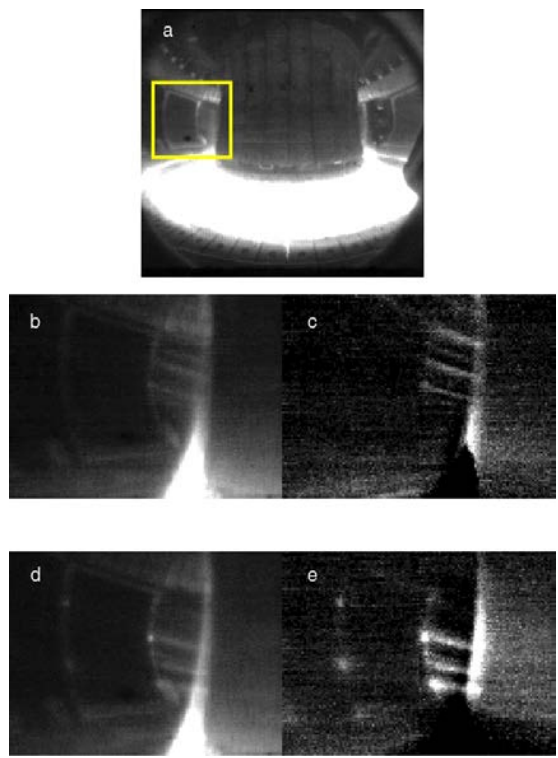

**Figure 2** Visible images obtained on ASDEX Upgrade. a) full view of the plasma with the region of fast acquisition shown as a box. b) image obtained during the start of the rise time of the target $D_\alpha$ due to an ELM and d) an image obtained 50 μs later during the same ELM. c) and e) are background subtracted versions of b and d respectively.

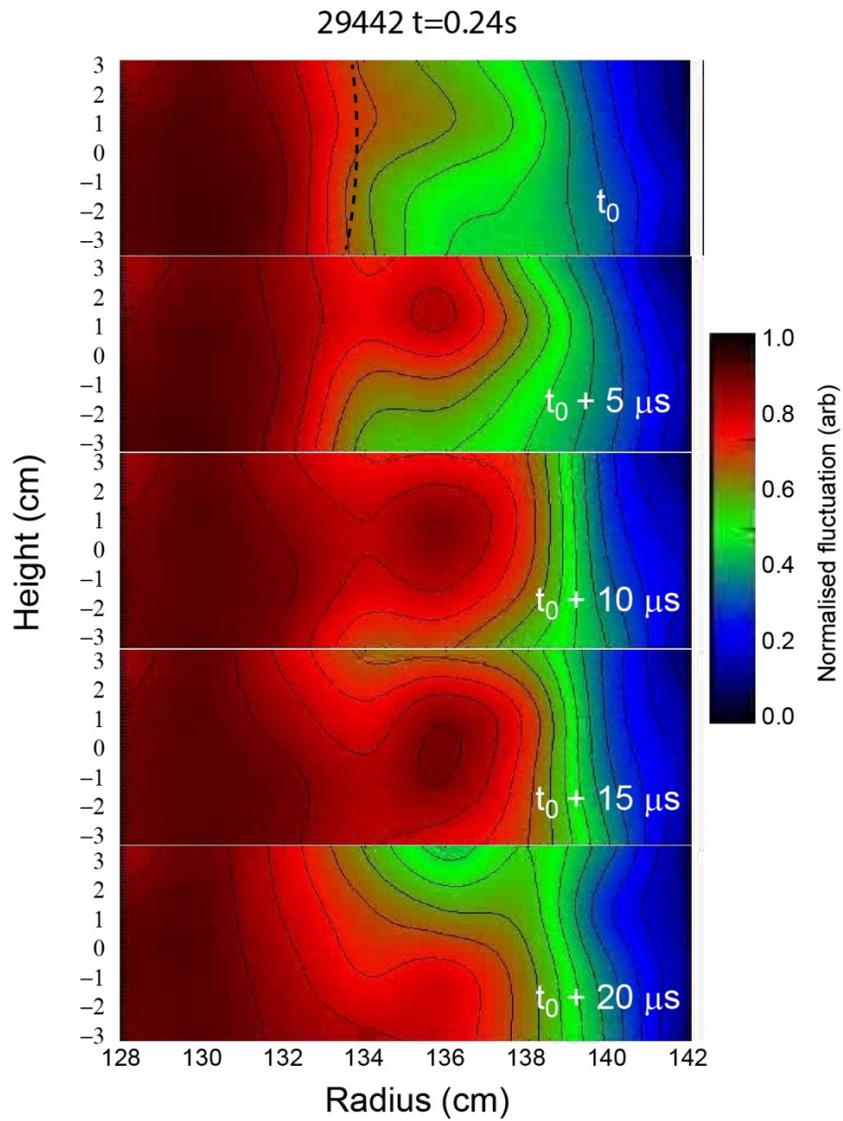

**Figure 3** Turbulence imaging of the plasma edge during an ELM crash on MAST using a Beam Emission Spectroscopy system. The frames shown are separated in time by 5µs relative to a notional crash time ($t_0$). The dotted curve in the first frame shows the location of the last closed flux surface.

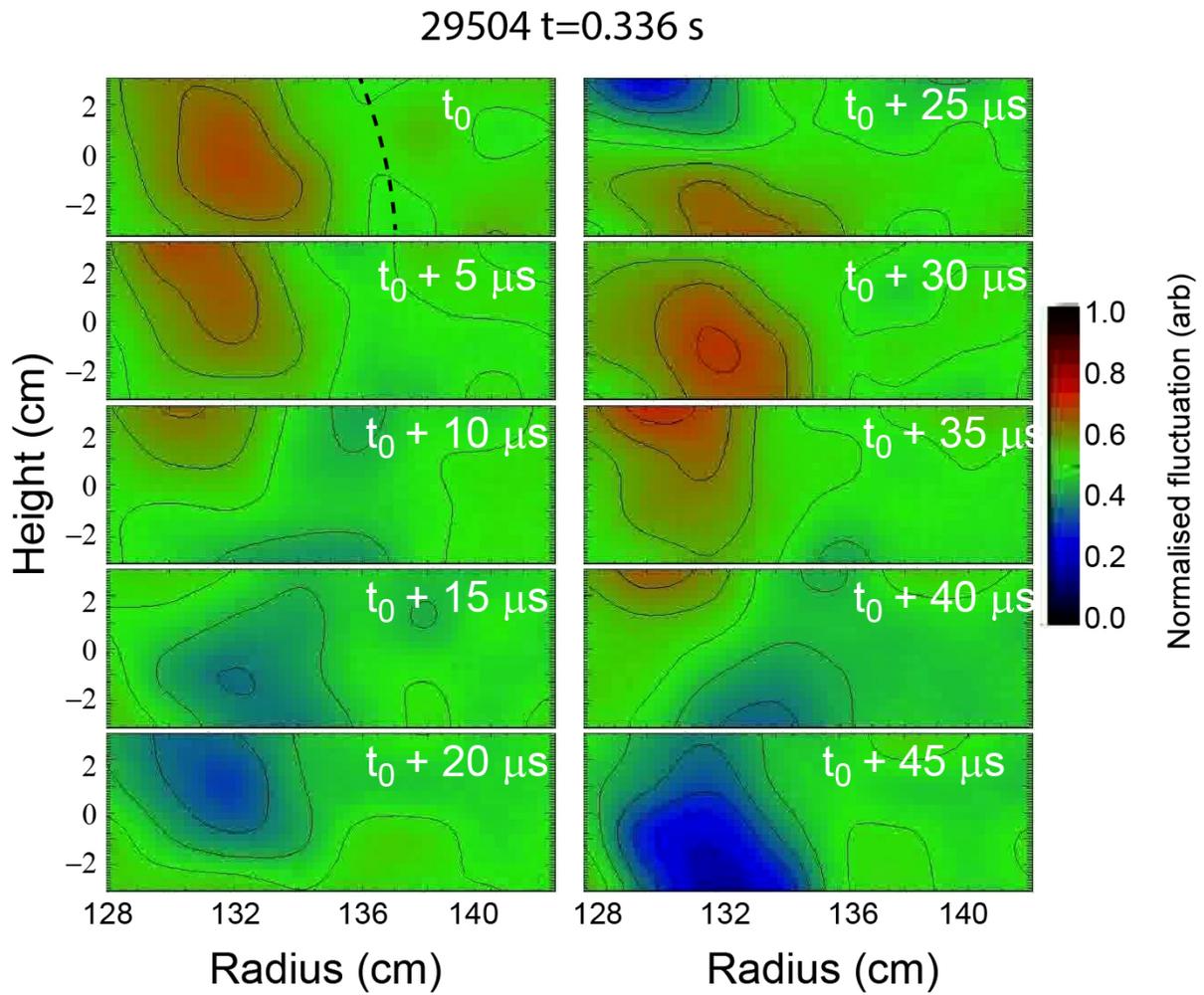

**Figure 4** Turbulence imaging of the plasma edge during the pre-cursor stage leading up to a type I ELM crash on MAST using a Beam Emission Spectroscopy system. The frames shown are separated in time by 5µs relative to a notional time ($t_0$), which is ~100 µs before the ELM crash. The dotted curve in the first frame shows the location of the last closed flux surface.

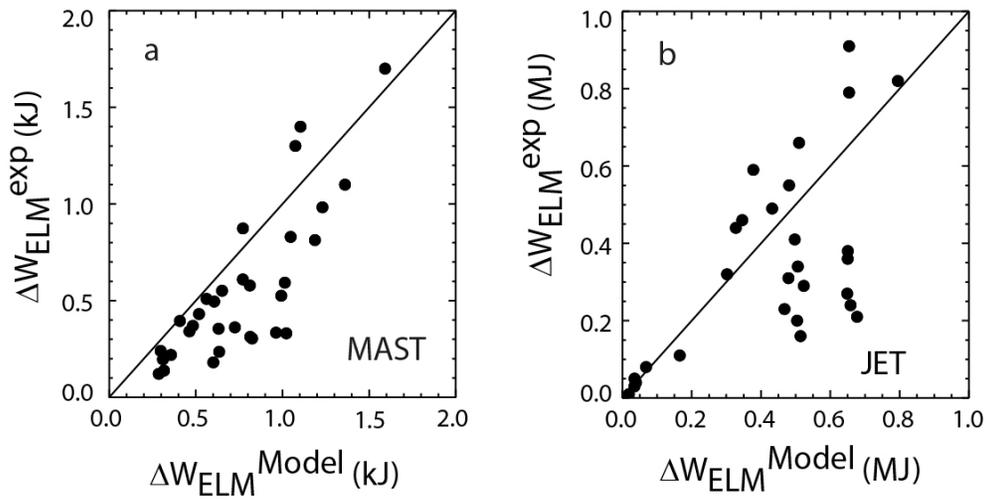

**Figure 5** Measured ELM energy loss ($\Delta W_{ELM}^{exp}$) versus the ELM energy loss calculated using a "leaky hosepipe" model ($\Delta W_{ELM}^{model}$) for a) MAST and b) JET.

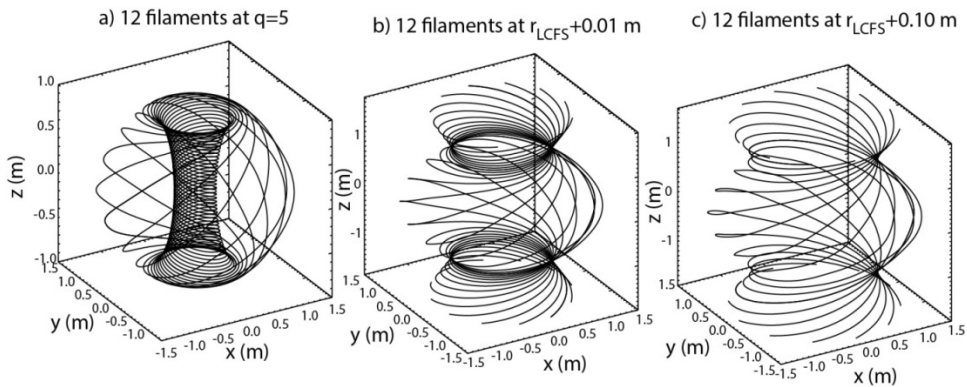

**Figure 6** Simulation of 12 equal toroidally spaced current carrying filaments located a) inside the LCFS on the q95=5 surface, b) 1cm and c) 10 cm outside the LCFS for a MAST like plasma in a connected double null magnetic configuration.

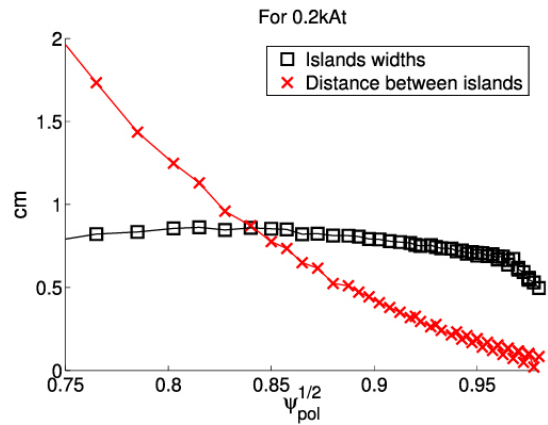

**Figure 7** Calculated island size (squares) and distance between islands (crosses) calculated in the vacuum approximation assuming 12 equally space filaments, each carrying 200 A, located at the q95=5 surface.

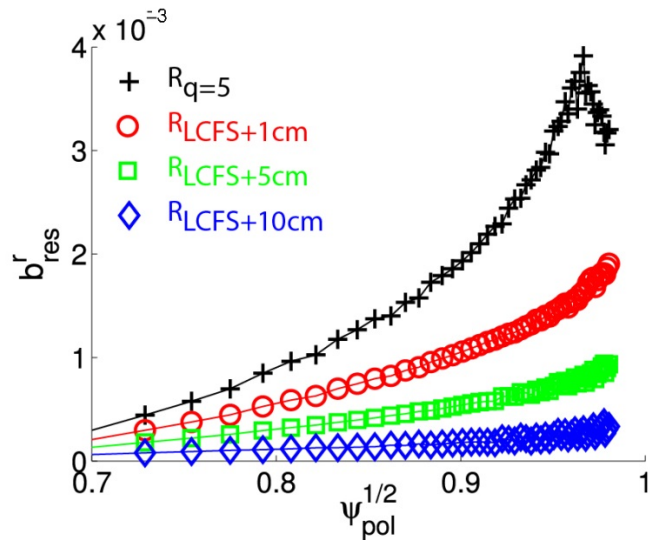

**Figure 8** Radial profile of the normalised resonant component of the applied field ($b^r_{res}$) for 12 equally space filaments each carrying 200 A located inside the LCFS at the $q_{95}$=5 surface (crosses) and a distance of 1 (circles) 5 (squares) and 10 cm (diamonds) outside the LCFS.

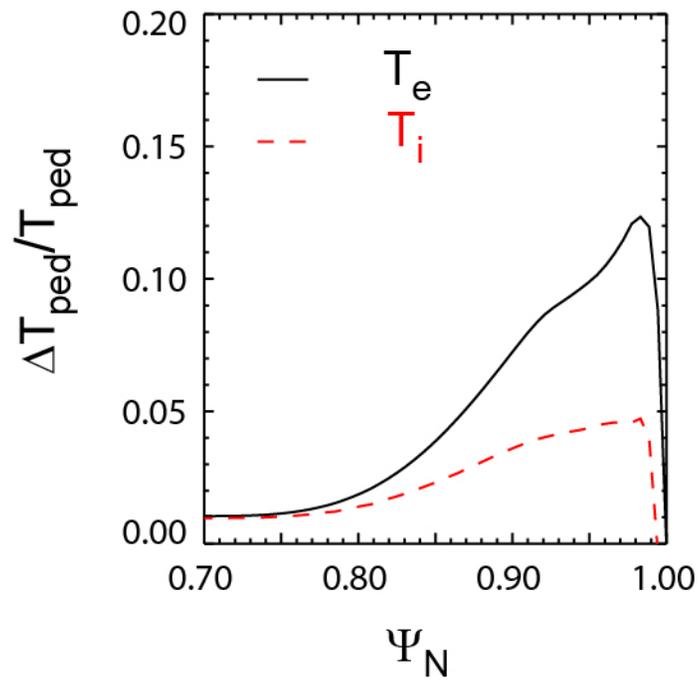

**Figure 9** JOREK simulations of the changes in the electron (solid) and ion temperature profiles due to an ELM in JET normalised to the temperature before the ELM.